\documentclass[preprint,showpacs,preprintnumbers,superscriptaddress,amsmath,amssymb]{revtex4}

\usepackage{epsfig}
\usepackage{dcolumn}
\usepackage{bm}

\begin{document}


\title{The extended empirical process test for non-Gaussianity in the CMB, with an application to non-Gaussian inflationary models}

\author{Frode K. Hansen}
\email{frodekh@roma2.infn.it}
\affiliation{Dipartimento di Fisica, Universit\`a di Roma `Tor Vergata', Via della Ricerca Scientifica 1, I-00133 Roma, Italy}
\author{Domenico Marinucci}
\email{marinucc@mat.uniroma2.it}
\affiliation{Dipartimento di Matematica, Universit\`a di Roma `Tor Vergata', Via della Ricerca Scientifica 1, I-00133 Roma, Italy}
\author{Nicola Vittorio}
\affiliation{Dipartimento di Fisica, Universit\`a di Roma `Tor Vergata', Via della Ricerca Scientifica 1, I-00133 Roma, Italy}
\affiliation{INFN, Sezione di Roma 2, Via della Ricerca Scientifica 1, I-00133 Roma, Italy}

\date{\today}

\begin{abstract}
In (Hansen et al. 2002) we presented a new approach for measuring non-Gaussianity of the Cosmic Microwave Background (CMB) anisotropy pattern, based on the multivariate empirical distribution function of the spherical harmonics $a_{\ell m}$ of a CMB map. The present paper builds upon the same ideas and proposes several improvements and extensions. More precisely, we exploit the additional information on the random phases of the $a_{\ell m}$ to provide further tests based on the empirical distribution function. Also we take advantage of the effect of rotations in improving the power of our procedures. The suggested tests are implemented on physically motivated models of non-Gaussian fields; Monte-Carlo simulations suggest that this approach may be very promising in the analysis of non-Gaussianity generated by non-standard models of inflation. We address also some experimentally meaningful situations, such as the presence of instrumental noise and a galactic cut in the map.
\end{abstract}

\maketitle

\section{Introduction}
\label{sect:intro}
Much information is expected from ongoing and future experiments aimed at measuring the CMB radiation. In this framework, a considerable amount of attention has been drawn in the recent literature by Gaussianity tests on CMB data. Such tests have been motivated by the prediction of the so-called theory of inflation, currently the most popular theory of the very early universe. Standard inflation predicts that the temperature fluctuations of the CMB, should be (very close to) Gaussian distributed (see the reviews \cite{narlikar,watson}). On the other hand more recent, non-standard models of inflation predict small non-Gaussianities; detecting such effects would be important for understanding the physics of the primordial epochs \cite{nongi1,nongi2,nongi3,sabino,nongi5,nongi6}. More radical deviations from Gaussianity are predicted by models where topological defects are induced by phase transitions (See Ref.\cite{strings} for a review). Detecting non-Gaussianities is also relevant as a tool for controlling systematic effects in CMB data. Among the many procedures advocated in the literature, many are based upon topological properties of spherical Gaussian fields \cite{novikov,gott,eriksen,heavens,vittorio,dore,barreiro2}; others are based on harmonic space properties \cite{phillips,komatsu,win,hu,kunz,barreiro1}  For empirical results, see \cite{boom,polenta,cobeng1,cobeng2}.\\

In a recent paper \cite{paper1} we proposed a new procedure to detect non-Gaussianity in harmonic space using so-called empirical processes \cite{shorack,mar}. More precisely, let $T(\theta ,\varphi )$ denote the CMB
fluctuations field, which we assume, as always, to be homogeneous and
isotropic, for $0<\theta \leq \pi ,$ $0<\varphi \leq 2\pi $. Assuming that $%
T(\theta ,\varphi )$ has zero mean and finite second moments, it is
well-known that the following spectral representation holds: 
\begin{equation*}
T(\theta ,\varphi )=\sum_{\ell=1}^{\infty }\sum_{m=-\ell}^{\ell}a_{\ell m}Y_{\ell m}(\theta
,\varphi )\text{ ,}
\end{equation*}%
where $Y_{\ell m}(\theta ,\varphi )$ denotes the spherical harmonics. The random
coefficients (amplitudes) $\left\{ a_{\ell m}\right\} $ have zero-mean with
variance $\langle|a_{\ell m}|^{2}\rangle=C_{\ell}$. They are uncorrelated over $\ell$ and $|m|$: $\langle a_{\ell m}a_{\ell^{\prime }m^{\prime }}^{\ast }\rangle=C_{\ell}\delta
_{\ell}^{\ell^{\prime }}\delta _{m}^{m^{\prime }}$, and $a_{\ell,m}=a_{\ell,-m}^{%
\ast }$. The sequence $\left\{ C_{\ell}\right\} $ denotes the angular power
spectrum of the random field and the asterisk complex conjugation.
Furthermore, if $T(\theta ,\varphi )$ is Gaussian, the $\left\{
a_{\ell m}\right\} $ have a complex Gaussian distribution. Upon observing $%
T(\theta ,\varphi )$ on the full sky, the random coefficients can be obtained through the
inversion formula%
\begin{equation}
a_{\ell m}=\int_{-\pi }^{\pi }\int_{0}^{\pi }T(\theta ,\varphi )Y_{\ell m}^{\ast
}(\theta ,\varphi )\sin \theta d\theta d\varphi \text{ , }m=0,\pm 1,...,\pm l%
\text{ },\text{ }l=1,2,...\text{ .}  \label{alm}
\end{equation}%

Our idea in that paper was to study the empirical distribution function for the $\left\{a_{\ell m}\right\} ,$ and to use these results to implement tests for
non-Gaussianity in harmonic space, in the presence of an unknown angular power spectrum. Although the Monte Carlo evidence produced in \cite{paper1} was rather encouraging, several important issues were left for further research. Indeed, it must be noted that although the joint distribution of the spherical harmonic coefficients is invariant under rotations (under Gaussianity), their numerical value is not; as a result, the sample outcome of the test in \cite{paper1} is not rotationally invariant. In this paper this remark is exploited to implement an extended, more powerful version of the empirical process procedure, based upon the combination of results from several different choices of coordinate systems. Furthermore, empirical process ideas are also implemented in four different setups: whereas in \cite{paper1} we focused only on the squared spherical harmonic coefficients, here we consider also the real and imaginary parts as separate random variables, which allows us to gain further efficiency improvements. More important, we consider the empirical process on the random phases of the Gaussian field. This approach is especially convenient, because under Gaussianity not only the random phases are model independent (compare \cite{naselsky}) but the Smirnov uniformization (see below) is exact, i.e. no bias arises. Moreover the joint information on the random phases and the random amplitudes uniquely identifies the distribution of the field, under Gaussianity. We address also some experimentally meaningful situations, such as the presence of instrumental noise and a galactic cut in the map; concerning the latter, we advocate a solution which may have some independent interest for other harmonic space methods of statistical inference. Our tests are implemented on  models of non-Gaussianity based upon ongoing research on non-standard inflationary models.\\

The plan of the paper is as follows: in section (\ref{sect:method}) we review the empirical process method; the extensions provided in this paper are presented in (\ref{sect:improvements}). Section (\ref{sect:results}) presents the results from Monte-Carlo simulations on 100 realizations of a non-standard inflationary model, with varying levels of non-Gaussianity. Comments and directions for further research are collected in the final section (\ref{sect:disc}).

\section{The empirical process method}
\label{sect:method}
The details of the empirical process approach to detect non-Gaussianity in the CMB were given in \cite{paper1}. In short, the method consists of a family of tests which focus on the total distribution of $a_{\ell m}$ and checks for dependencies between $k$ $\ell$-rows. The first step is to transform the spherical harmonic coefficients into variables $u_{\ell m}$ which have an approximate uniform distribution between $0$ and $1$, given that the $a_{\ell m}$ were initially Gaussian distributed. This is done using the Smirnov transformation, defined as
\begin{equation*}
u_{\ell 0}=\Phi _{1}\left(\frac{|a_{\ell 0}|^{2}}{\hat C_{\ell}}\right)\text{ , }u_{\ell m}=\Phi _{2}\left(\frac{%
2|a_{\ell m}|^{2}}{\hat C_{\ell}}\right)\text{ , }m=1,2,...,l,\text{ }l=1,2,...L\text{ ,}
\end{equation*}%
where $\Phi_n$ is the cumulative distribution function of a $\chi^2$ with $n$ degrees of freedom and $\hat C_\ell$ are the power spectrum coefficients estimated from the data. The error introduced by using estimated $\hat C_\ell$ instead of the real underlying $C_\ell$ is dealt with using a bias-subtraction, as described in \cite{paper1}.\\

Then the joint empirical distribution function for row $\ell$ is formed,
\begin{equation*}
\widehat{F}_{\ell...\ell+\Delta_{\ell,k-1}}(\alpha _{1},...,\alpha _{k})=\frac{1}{(\ell+1)}%
\sum_{m=0}^{\ell}\left\{ \underline{\mathbf{1}}(\widehat{u}_{\ell m}\leq \alpha
_{1})\prod_{i=2}^{k}\underline{\mathbf{1}}(\widehat{u}_{\ell+\Delta_{\ell,i-1},m+\Delta _{mi}}\leq \alpha
_{i})\right\} \text{ , }\Delta _{mi}\geq 0\text{ ,}
\end{equation*}
where $\Delta_{\ell,i}$ determines the spacing between the rows for which the dependencies are tested and $\Delta_{mi}$ denotes the difference in $m$ for row $i$. The parameters $\alpha_i$ run over the interval $[0,1]$. The empirical process is expressed using the centered and rescaled $\widehat{F}_{\ell...\ell+\Delta_{\ell,k-1}}$ given as
\begin{equation*}
\widehat{G}_{\ell...\ell+\Delta_{\ell,k-1}}(\alpha _{1},...,\alpha _{k})=\sqrt{(\ell+1)}%
\left\{ \widehat{F}_{\ell...\ell+\Delta_{\ell,k-1}}(\alpha _{1},...,\alpha
_{k})-\prod_{i=1}^{k}\alpha _{i}\right\} \text{ .}
\end{equation*}
The intuition behind this procedure is as follows: if the $a_{\ell m}$s are Gaussian, $\widehat{G}$ converges to a well-defined limiting process, whose distribution can be readily tabulated. On the other hand, for non-Gaussian $a_{\ell m}$s $\{\widehat{F}(\alpha_{1},...,\alpha_{k})-\prod_{i=1}^{k}\alpha_{i}\}$ and thereby $\widehat{G}$ will take `high' values over some parts of $\alpha$-space. Thus, the analysis of some appropriate functional of $\widehat{G}$ can be used to detect non-Gaussianity. To combine the information over all multipoles into one statistic, we define
 \begin{equation*}
\widehat{K}_{L}(\alpha _{1},...,\alpha _{k},r)=\frac{1}{\sqrt{L-\Delta
_{\ell,k-1}}}\sum_{\ell=1}^{[(L-\Delta _{k-1})r]}\widehat{G}_{\ell,...,\ell+\Delta
_{\ell,k-1}}(\alpha _{1},...,\alpha _{k})\text{ ,}
\end{equation*}
where $L$ is the highest multipole where the data is signal dominated.

The method can then be summarized as follows: the distribution of  $\mathrm{sup}|\widehat{K}_{L}|$ is found using Monte-Carlo simulations of Gaussian distributed $a_{\ell m}$. Then, for a given observed set of $a_{\ell m}$, the value $k_\mathrm{max}=\mathrm{sup}|\widehat{K}_{L}|$ is found and compared to the distribution obtained from Monte-Carlo. The consistency of the data with a Gaussian distribution can then be estimated to any suitable $\sigma$-level. 

\section{Improvements of the empirical process method}
\label{sect:improvements}
The process described in the previous section can be improved in several ways. One important thing to notice is that if the $a_{\ell m}$ are non-Gaussian distributed, their joint distribution will change under a rotation of the sky, and therefore it will not be invariant with respect to the choice of coordinate system. In spherical harmonic space, the rotation of the sky with the Euler angles $(\Phi_1,\Theta,\Phi_2)$ can be written as,
\begin{equation}
\label{eq:almr}
a^R_{\ell m}=\sum_{m'}D_{mm'}^\ell(\Phi_1,\Theta,\Phi_2) a_{\ell m'},
\end{equation}
where the rotation matrices $D_{mm'}^\ell(\Phi_1,\Theta,\Phi_2)$ are described in Appendix (\ref{app:rot}). This new sets of $a_{\ell m}$ will yield a different (although not independent) value of $k_\mathrm{max}$; averaging the $k_\mathrm{max}$ obtained from the empirical process over several rotations may thus increase the probability for detecting non-Gaussianity in the data. In Table \ref{tab:arot}, we indeed prove that this effect is noticeable, considering a non-Gaussian inflationary model with $f_{NL}=300$. To obtain this data, we used 100 Gaussian Monte-Carlo simulations to find the distribution of $k_\mathrm{max}$. Then we used 100 realizations of the non-Gaussian model and for each realization we averaged the $k_\mathrm{max}$ value obtained over 45 rotation in the $\Theta$-direction. The results are shown using the process to test Gaussianity and dependency between 1, 2 and 3 $\ell$-rows. We define the $1\sigma$, $2\sigma$ and $3\sigma$ detection levels as the $k_\mathrm{max}$ values over which we had $32\%$, $5\%$ and $1\%$ of the hits in the Gaussian simulations respectively.\\

\begin{center}
\begin{table}
\caption{INFLATIONARY MODEL, $f_{NL}=300$, test A}
\label{tab:arot}
\begin{ruledtabular}
\begin{tabular}{|l|l|l|l|l|l|l|l|}
number of rows: & 1 & 2 & 3 & 45 $\Theta$-rotations: & 1 &  2 & 3 \\
\hline
1$\sigma$ & 28\% & 35\% & 44\% & & 32\% & 41\% & 52\% \\ 
2$\sigma$ & 5\% & 7\% & 14\% & & 4\% & 22\% & 23\%  \\ 
3$\sigma$ & 3\% & 4\% & 6\% & & 1\% & 8\% & 10\%  
\end{tabular}
\end{ruledtabular}
\end{table}
\end{center}

As seen from equation (\ref{eq:almr}), a rotation in the $\Phi$-direction will only change the phase of the $a_{\ell m}$ for which the empirical process as defined above is insensitive. The reason is that we define the $u_{\ell m}$ using the modulus of the $a_{\ell m}$. But by defining a complex $u_{\ell m}$ for which the real and imaginary parts are just the uniformized real and imaginary parts of the $a_{\ell m}$, the phase information can be kept, and the test will be sensitive also to rotations in the $\Phi$ direction. The new $u_{\ell m}$ can be obtained by a similar Smirnov transformation,
\begin{equation*}
u^r_{\ell m}=G\left(\frac{a^r_{\ell m}}{\sqrt{\hat C_{\ell}}}\right)\text{ , }u_{\ell m}^i=G\left(\frac{%
a^i_{\ell m}}{\sqrt{\hat C_{\ell}}}\right)\text{ , }m=1,2,...,l,\text{ }l=1,2,...L\text{ ,}
\end{equation*}%
where $G$ is the cumulative distribution function of a Gaussian variable with zero mean and unit variance and the $r$ and $i$ indices are used to indicate real and imaginary parts.\\

With these new $u_{\ell m}$ the standard empirical process can be applied, now also allowing the check for dependencies between real and imaginary parts. We will in this paper arrange the $u_{\ell m}$ in two ways. First we will double the number of rows, putting the imaginary rows in between the real rows, then we apply the standard empirical process using $\Delta\ell=\Delta m=250$ for $L=500$ as in \cite{paper1} (from now on we will label this the C test). Otherwise we will just double the length of the $\ell$-rows, putting the imaginary parts on the negative $m$ side. Then we apply the standard empirical process still with $\Delta\ell=\Delta m=250$ for $L=500$ (called the B test). In both ways we check dependencies between real and imaginary parts. In Table \ref{tab:3tst} we show results from these two new tests applied to 100 realizations of the $f_{NL}=300$ model used above (to be compared with the results of test A without rotations in Table \ref{tab:arot}). Even if the number of detections in each test vary, the specific realizations detected by each of the tests also vary. This suggests that by combining the results from all tests, the number of detections can be increased. We suggest to make the average $k_\mathrm{max}$ over all 3 tests as a new statistic. The average has to be weighted with the mean $\bar k_\mathrm{max}$ for each test, obtained from Monte-Carlo. Thus the new statistic would be:
\begin{equation*}
k_\mathrm{max}=k_\mathrm{max}^A+\frac{\bar k_\mathrm{max}^A}{\bar k_\mathrm{max}^B}k_\mathrm{max}^B+\frac{\bar k_\mathrm{max}^A}{\bar k_\mathrm{max}^C}k_\mathrm{max}^C,
\end{equation*}
where the letters indicate test A, B and C.\\

\begin{center}
\begin{table}
\caption{INFLATIONARY MODEL, $f_{NL}=300$, test B,C and A+B+C}
\label{tab:3tst}
\begin{ruledtabular}
\begin{tabular}{|l|l|l|l|l|l|l|l|l|l|l|l|}
rows(B):: & 1 & 2 & 3 & rows(C): & 1 &  2 & 3 & rows(A+B+C): & 1 & 2 & 3 \\
\hline
1$\sigma$ & 31\% & 32\% & 41\% & & 30\% & 31\% & 38\% & & 31\% & 38\% & 50\% \\ 
2$\sigma$ & 8\% & 8\% & 13\% & & 4\% & 7\% & 12\% & & 6\% & 6\% & 17\% \\ 
3$\sigma$ & 1\% & 2\% & 2\% & & 1\% & 1\% & 2\%  & & 1\% & 1\% & 2\%
\end{tabular}
\end{ruledtabular}
\end{table}
\end{center}

As we showed above, the old test can be improved significantly using rotations in the $\Theta$-directions. The two new tests will also be sensitive to rotations in the $\Phi$-directions, as these rotations only change the phases of the $a_{\ell m}$. In Table \ref{tab:phirot} we show a similar increase in the number of detections using 30 rotations in the $\Phi$-rotation. We also show the total number of detections, combining test A averaged over 30 $\Theta$-rotations and test B and C averaged over 30 $\Phi$-rotations. Clearly the highest number of detections can be obtained combining all 3 tests with as many rotations as possible for each, and in this way testing all possible dependencies in spherical harmonic space.\\

Finally the same sort of ideas can be implemented by working directly on the random phases by taking the uniformized coefficients to be $u_{\ell m}=\arctan(a_{\ell m}^i/a_{\ell m}^r)/\pi$. From a theoretical point of view, this choice is particularly appealing because the uniformization is exact. Random phases as a test for non-Gaussianity were considered also by \cite{naselsky}; these authors however do not consider the empirical process approach. We label this test D and we report the related results in Table \ref{tab:phase}.

In practice, a relevant issue is the coupling between spherical harmonic coefficients introduced by partial sky coverage from a galactic cut. It turns out that this can be well controlled if the Monte-Carlo calibration of the $k_\mathrm{max}$ distribution are done with the same galactic cut. The introduction of the galactic cut has the effect of reducing the number of detections, but this can be significantly improved if the gap is refilled, using the two borders between the gap and the observed area as a mirror. This creates a discontinuity at the galactic equator, but improves the power of the procedure. In our test, using this refilling technique, we found no noticeable reduction in the number of detections with and without a galactic cut of $5^\circ\times2$ in the 100 realizations used in the simulations.

\begin{center}
\begin{table}
\caption{INFLATIONARY MODEL, $f_{NL}=300$, test B,C and A+B+C with 30 rotations}
\label{tab:phirot}
\begin{ruledtabular}
\begin{tabular}{|l|l|l|l|l|l|l|l|l|l|l|l|}
rows(B):: & 1 & 2 & 3 & rows(C): & 1 &  2 & 3 & rows(A+B+C): & 1 & 2 & 3 \\
\hline
1$\sigma$ & 33\% & 31\% & 67\% & & 28\% & 30\% & 50\% & & 30\% & 42\% & 75\% \\ 
2$\sigma$ & 8\% & 5\% & 29\% & & 6\% & 4\% & 18\% & & 8\% & 15\% & 32\% \\ 
3$\sigma$ & 2\% & 3\% & 15\% & & 1\% & 3\% & 16\%  & & 4\% & 10\% & 17\%
\end{tabular}
\end{ruledtabular}
\end{table}
\end{center}

\section{Detection of non-Gaussian inflationary models}
\label{sect:results}

In this section we will show the results of the 4 versions of the empirical process method applied to 100 realizations of a non-standard inflationary model. Indeed, as discussed in \cite{sabino}, one way of parameterizing the possible presence of non-Gaussianity in the primordial gravitational potential $\Phi$ is to represent it in the following way,
\begin{equation}
\Phi=\phi+f_{NL}(\phi^2-<\phi^2>),
\end{equation}
where $\phi$ is a zero mean Gaussian random field and $f_{NL}$ is a non-linear parameter which can be observationally constrained. Our empirical examples are based on this model, using for simplicity a pure Sachs-Wolfe angular power spectrum ($C_\ell\propto\ell(\ell+1)$). It should be noted that because $\phi$ is of the order $10^{-5}$, the quadratic term is five orders of magnitudes smaller than the linear one.\\

\begin{figure}[tbp]
\begin{center}
\leavevmode
\epsfig {file=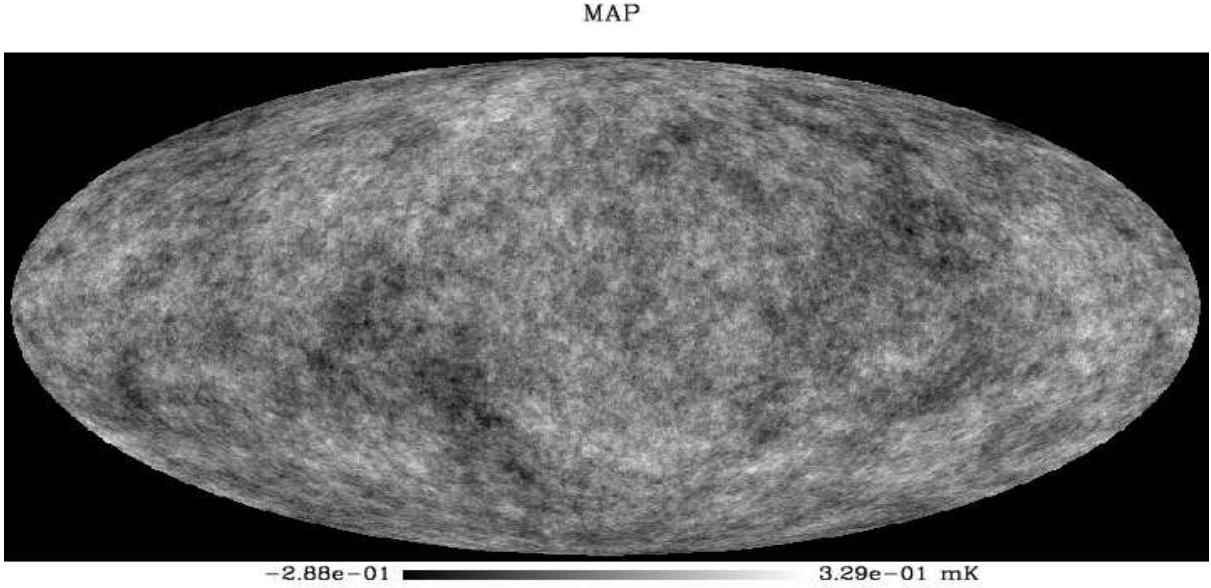,width=16cm,height=8cm}
\caption{Map with a non-Gaussian part $f_{NL}=100$.}
\label{fig:map100}
\end{center}
\end{figure}

\begin{figure}[tbp]
\begin{center}
\leavevmode
\epsfig {file=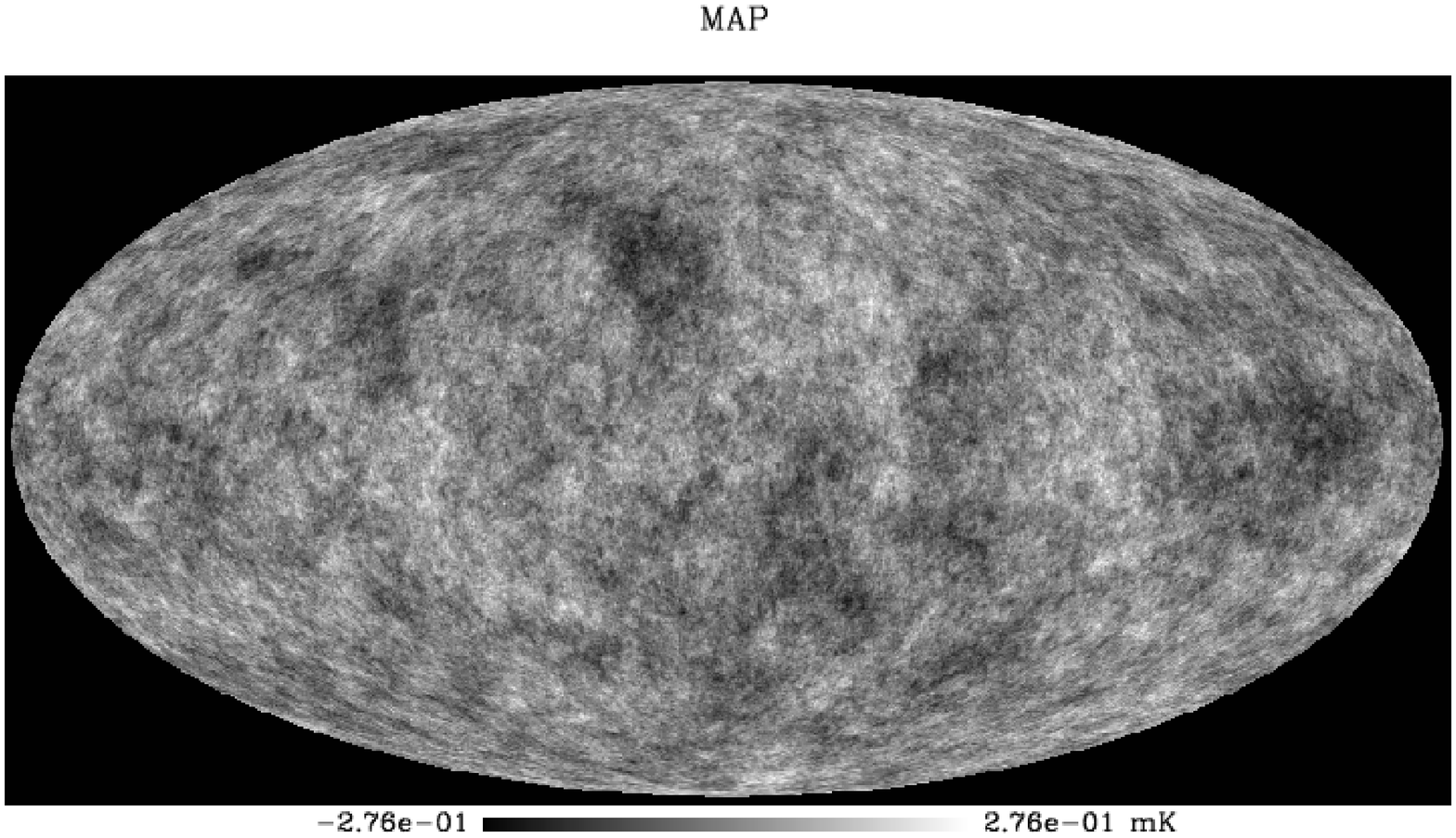,width=16cm,height=8cm}
\caption{Map with a non-Gaussian part $f_{NL}=300$.}
\label{fig:map300}
\end{center}
\end{figure}

\begin{figure}[tbp]
\begin{center}
\leavevmode
\epsfig {file=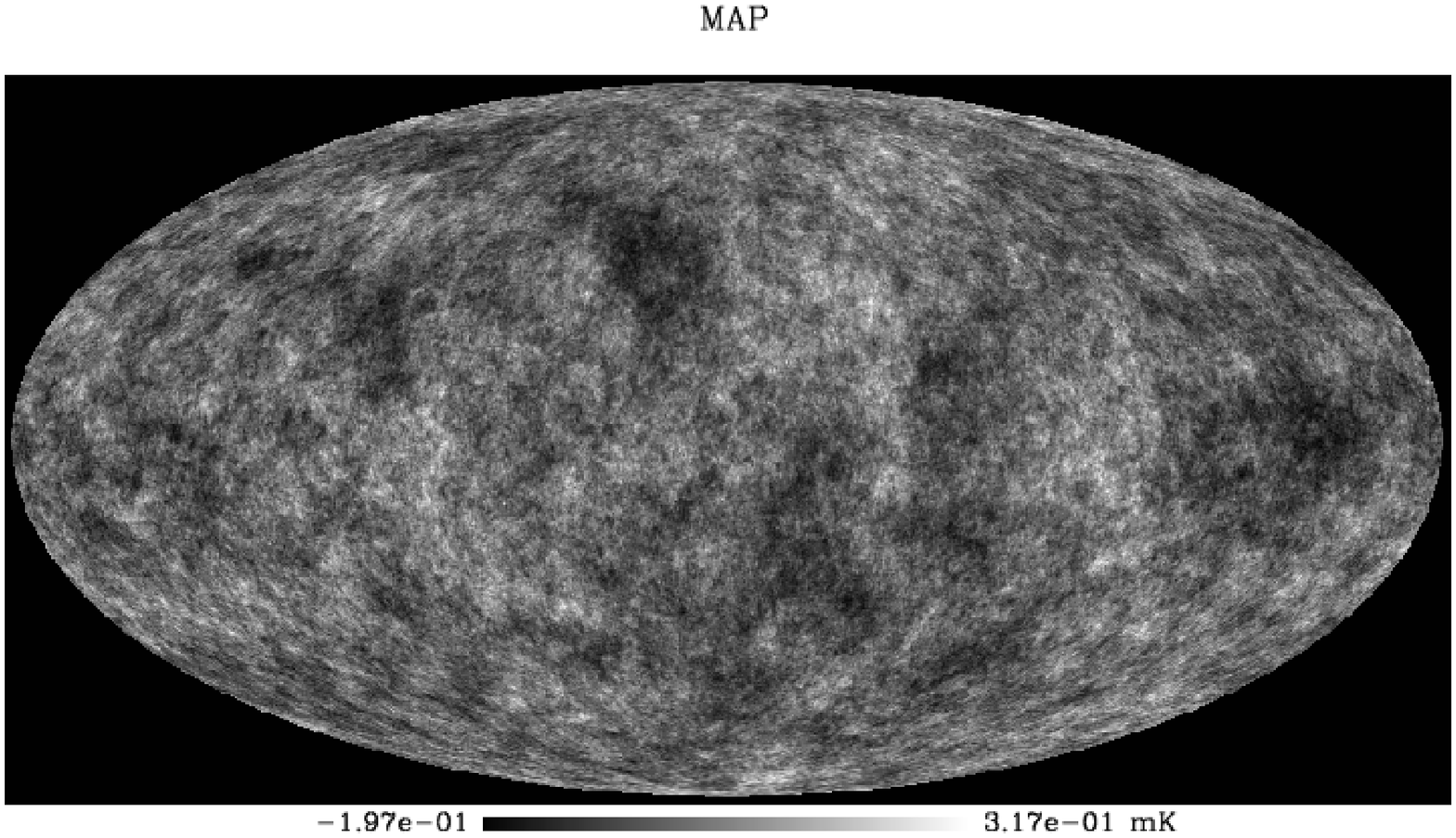,width=16cm,height=8cm}
\caption{Map with a non-Gaussian part $f_{NL}=1000$.}
\label{fig:map1000}
\end{center}
\end{figure}

\begin{figure}[tbp]
\begin{center}
\leavevmode
\epsfig {file=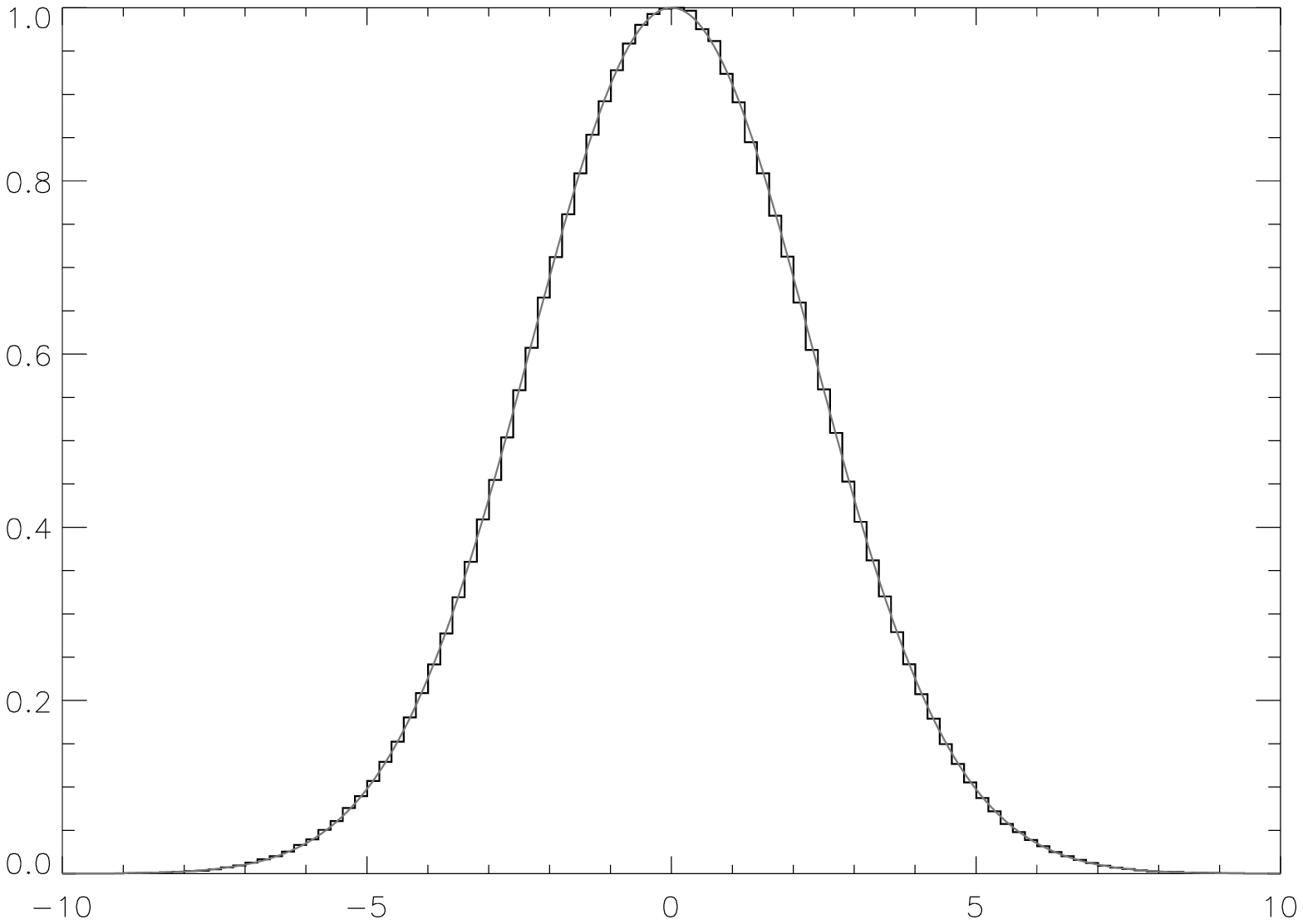,width=12cm,height=8cm}
\caption{Histogram of the map at $f_{NL}=100$. The shaded curve is a Gaussian fit.}
\label{fig:hist100}
\end{center}
\end{figure}

\begin{figure}[tbp]
\begin{center}
\leavevmode
\epsfig {file=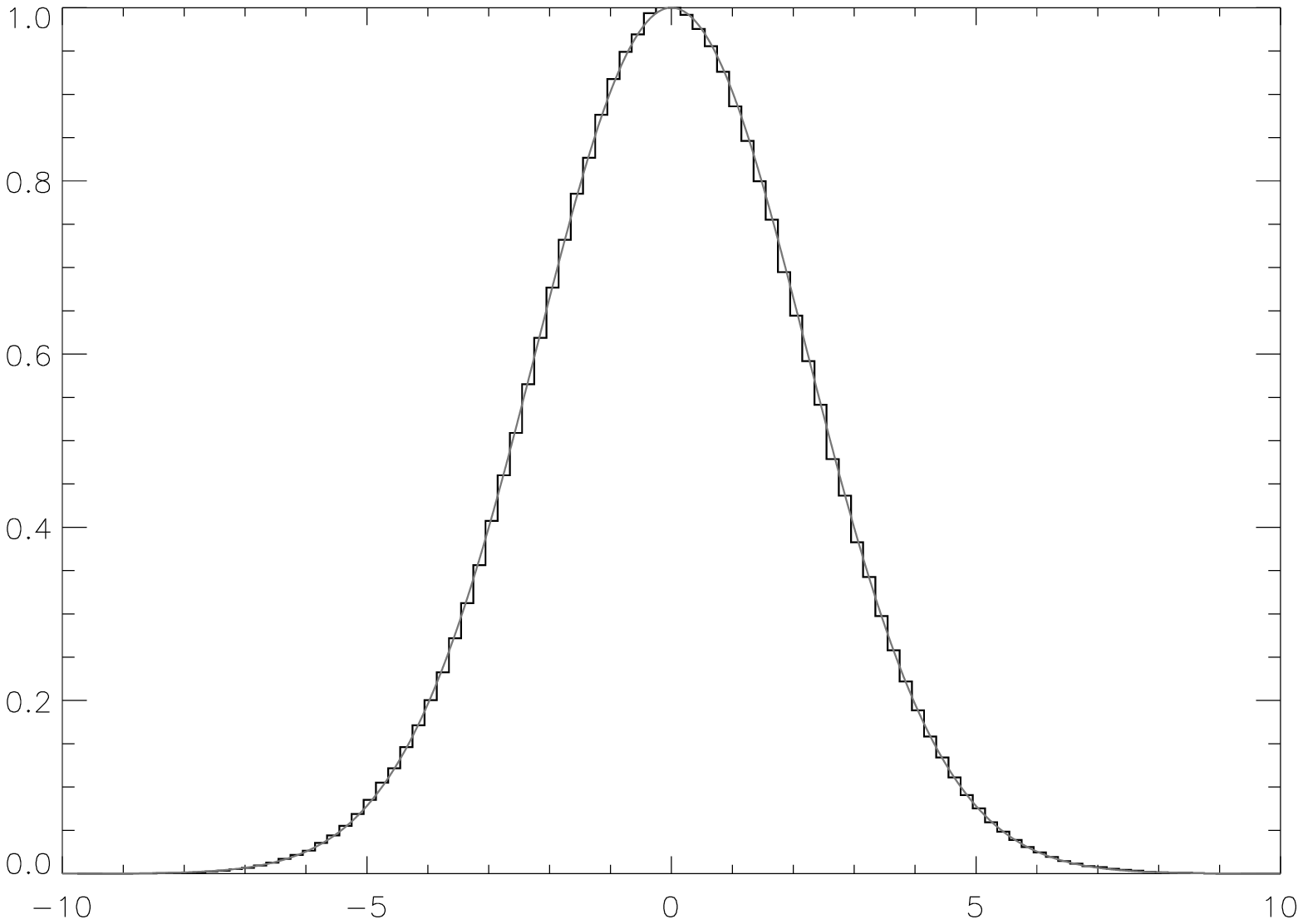,width=12cm,height=8cm}
\caption{Histogram of the map at $f_{NL}=300$. The shaded curve is a Gaussian fit.}
\label{fig:hist300}
\end{center}
\end{figure}

\begin{figure}[tbp]
\begin{center}
\leavevmode
\epsfig {file=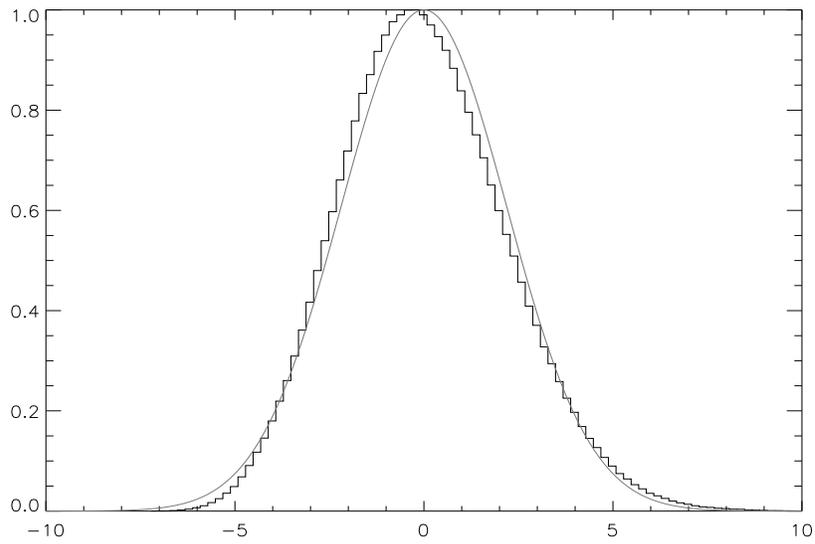,width=12cm,height=8cm}
\caption{Histogram of the map at $f_{NL}=1000$. The shaded curve is a Gaussian fit.}
\label{fig:hist1000}
\end{center}
\end{figure}

 We will use 3 levels of non-Gaussianity, $f_{NL}=100$, $f_{NL}=300$ and $f_{NL}=1000$ (a map for each $f_{NL}$ with its corresponding histogram is shown in figures \ref{fig:map100}-\ref{fig:hist1000}). On these realizations we put non-uniform noise corresponding to the noise-level as expected from the Planck LFI $100\mathrm{GHz}$ channel. We also include a $5^\circ\times2$ galactic cut. The Monte-Carlo for determination of the distribution of $k_\mathrm{max}$ will when not otherwise indicated be made using the same maps with $f_{NL}=0$.\\

The results of the tests are reported in Tables \ref{tab:split}-\ref{tab:phase}. We see that the 3-rows version of the test provides the best performance. Among the 4 tests, none appears to detect significantly more than others. Down to a level $f_{NL}=300$, more than half of the maps are detected at the $2\sigma$ level. However at $f_{NL}=100$, the detection level at $2\sigma$ has dropped to about $30\%$ at the best. Nevertheless, as the spectrum we used went only to multipole $\ell=500$ and that radiative transfer was not taken into account, we expect that these numbers can
improve significantly.

\begin{center} 
\begin{table}
\caption{TEST B}
\label{tab:split}
\begin{ruledtabular}
\begin{tabular}{|l|l|l|l|l|l|l|l|l|l|}
$f_{NL}$ &  & 100 &  &  & 300 &  &  & 1000 &  \\
\hline 
$rows$ & 1 & 2 & 3 & 1 & 2 & 3 & 1 & 2 & 3 \\ 
\hline
1 $\sigma $ & 42\% & 34\% & 47\% & 32\% & 37\% & 83\% & 28\% & 70\% & 100\%
\\ 
2 $\sigma $ & 8\% & 7\% & 13\% & 6\% & 10\% & 59\% & 10\% & 38\% & 98\% \\ 
3 $\sigma $ & 2\% & 3\% & 9\% & 1\% & 4\% & 52\% & 1\% & 30\% & 98\%%
\end{tabular}
\end{ruledtabular}
\end{table}
\end{center}

\begin{center} 
\begin{table}
\caption{TEST C}
\label{tab:dep}
\begin{ruledtabular}
\begin{tabular}{|l|l|l|l|l|l|l|l|l|l|}
$f_{NL}$ &  & 100 &  &  & 300 &  &  & 1000 &  \\ 
\hline
$rows$ & 1 & 2 & 3 & 1 & 2 & 3 & 1 & 2 & 3 \\ 
\hline
1 $\sigma $ & 32\% & 33\% & 57\% & 36\% & 43\% & 93\% & 38\% & 84\% & 100\%
\\ 
2 $\sigma $ & 11\% & 8\% & 12\% & 9\% & 15\% & 68\% & 22\% & 78\% & 99\% \\ 
3 $\sigma $ & 2\% & 2\% & 3\% & 3\% & 12\% & 45\% & 7\% & 69\% & 99\%%
\end{tabular}
\end{ruledtabular}
\end{table}
\end{center}

\begin{center} 
\begin{table}
\caption{TEST B+C}
\label{tab:depsplit}
\begin{ruledtabular}
\begin{tabular}{|l|l|l|l|l|l|l|l|l|l|}
$f_{NL}$ &  & 100 &  &  & 300 &  &  & 1000 &  \\ 
\hline
$rows$ & 1 & 2 & 3 & 1 & 2 & 3 & 1 & 2 & 3 \\ 
\hline
1 $\sigma $ & 40\% & 34\% & 50\% & 36\% & 41\% & 95\% & 40\% & 87\% & 100\%
\\ 
2 $\sigma $ & 5\% & 8\% & 29\% & 6\% & 16\% & 87\% & 13\% & 77\% & 99\% \\ 
3 $\sigma $ & 2\% & 6\% & 10\% & 1\% & 11\% & 68\% & 2\% & 70\% & 99\%%
\end{tabular}
\end{ruledtabular}
\end{table}
\end{center}

\begin{center} 
\begin{table}
\caption{TEST D}
\label{tab:phase}
\begin{ruledtabular}
\begin{tabular}{|l|l|l|l|l|l|l|l|l|l|}
$f_{NL}$ &  & 100 &  &  & 300 &  &  & 1000 &  \\ 
\hline
$rows$ & 1 & 2 & 3 & 1 & 2 & 3 & 1 & 2 & 3 \\ 
\hline
1 $\sigma $ & 35\% & 37\% & 45\% & 39\% & 40\% & 83\% & 44\% & 47\% & 97\%
\\ 
2 $\sigma $ & 9\% & 4\% & 19\% & 16\% & 5\% & 72\% & 30\% & 17\% & 97\% \\ 
3 $\sigma $ & 0\% & 0\% & 7\% & 2\% & 1\% & 57\% & 9\% & 11\% & 97\%%
\end{tabular}
\end{ruledtabular}
\end{table}
\end{center}

\section{Comments and conclusions}
\label{sect:disc}

In \cite{paper1} we argued that the empirical process method enjoyed a number of advantages over existing methods, such as being completely model independent, allowing for a rigorous statistical analysis, detecting the location of non-Gaussianity in harmonic space, being easily amendable to extensions and generalizations. In this paper the potentials of this approach are further investigated. Three further classes of tests in the same spirit as in the previous paper are proposed. Changes of coordinates are implemented to make the procedure rotationally invariant and more powerful against non-Gaussian alternatives. We have also considered experimentally meaningful situations such as the presence of gaps and noise. Finally, we have tested the power of these procedures against physically reasonable models such as those expected by non-standard inflation; we show that for the model considered, significant detections of non-Gaussianities are achieved for non linear coupling parameters down to about 100. Overall we believe that empirical process methods emerge as valuable tools for further studies on the non-Gaussianity of the CMB.

\begin{acknowledgments}

We acknowledge the use of the Healpix package\footnote{http://www.eso.org/science/healpix/}. We are very grateful to Sabino Matarrese and Michele Liguori for supplying us with non-Gaussian maps. We are also very grateful to Amadeo Balbi, Paolo Natoli and Paolo Cabella for very useful discussions. FKH acknowledges financial support from the CMBNET Research Training Network.

\end{acknowledgments}

\appendix

\section{Rotation Matrices}
\label{app:rot}
A spherical function $T(\theta,\phi)$ is rotated by the operator $\hat
D(\alpha\beta\gamma)$ where $\alpha\beta\gamma$ are the three Euler
angles for rotations \cite{risbo} and the inverse rotation is
$\hat D(-\gamma-\beta-\alpha)$. For the spherical harmonic functions,
this operator takes the form,
\begin{equation}
Y_{\ell m}(\theta',\phi')=\sum_{m'=-\ell}^\ell
D_{m'm}^\ell(\alpha\beta\gamma)Y_{\ell m'}(\theta,\phi),
\end{equation}
where $D_{m'm}^\ell$ has the form
\begin{equation}
D_{m'm}^\ell(\alpha\beta\gamma)=e^{im'\alpha}d_{m'm}^\ell(\beta)e^{im\gamma}.
\end{equation}
Here $d_{m'm}^\ell(\beta)$ is a real coefficient with the following
property:
\begin{equation}
d_{m'm}^\ell(\beta)=d_{mm'}^\ell(-\beta).
\end{equation}
The D-functions also have the following property:
\begin{equation}
D_{m'm}^\ell(\alpha\beta\gamma)=\sum_{m''}D_{m'm''}^\ell(\alpha_2\beta_2\gamma_2)D_{m''m}^\ell(\alpha_1\beta_1\gamma_1),
\end{equation}
where $(\alpha\beta\gamma)$ is the result of the two consecutive
rotations $(\alpha_1\beta_1\gamma_1)$ and
$(\alpha_2\beta_2\gamma_2)$.

The complex conjugate of the rotation matrices can be written as
\begin{equation}
D^{\ell*}_{mm'}=(-1)^{m+m'}D^\ell_{(-m)(-m')}.
\end{equation}

\end{document}